\documentclass[twocolumn,secnumarabic,amssymb, nobibnotes, aps, prd]{revtex4-2}

\usepackage{lipsum}
\usepackage[separate-uncertainty = true]{siunitx}
\usepackage{graphicx}
\DeclareSIUnit{\degree}{^\circ}
\begin{document}

\title{Simulation of the impact of an additional  corrugated structure impedance on the bursting dynamics in an electron storage ring}

\author{S.~Maier}%
\email{sebastian.maier@kit.edu}
\author{M.~Brosi}
\thanks{Now at MAX IV, Lund, Sweden}
\author{A.~Mochihashi}
\author{M.~J.~Nasse}
\author{M.~Schwarz}
\author{A.-S.~Müller}
\affiliation{Karlsruhe Institute of Technology, Hermann-von-Helmholtz-Platz 1, 76344 Eggenstein-Leopoldshafen, Germany}
\date{\today}%
\begin{abstract}
In the case of single-digit picosecond bunch length, synchrotron light sources produce intense coherent radiation up to the \si{\tera\hertz} range. The reduction of the bunch length by lowering the momentum compaction factor (low-$\alpha$) gives rise to the micro-bunching instability, which is on one hand a crucial roadblock in the X-ray range during to the resulting effective bunch lengthening but on the other hand also an opportunity for the generation of intense \si{\tera\hertz} radiation if it can be controlled appropriately. In the KIT storage ring KARA (Karlsruhe Research Accelerator), two parallel plates with periodic rectangular corrugations are planned to be installed as a proof of principle in an electron storage ring. These plates create an additional longitudinal impedance based on their geometry, which can affect the beam dynamics. The resulting impedance manipulation will be used to study and control the longitudinal electron beam dynamics and the emitted coherent synchrotron radiation (CSR). This paper presents the results of systematic studies in simulation of the impact of additional corrugated plate impedances on the longitudinal beam dynamics using the example of the KARA storage ring. If the periodicity of the wake function of the corrugated plates matches the size of the substructures in the longitudinal bunch profile, the instability threshold can be effectively manipulated. This extends intense  \si{\tera\hertz} radiation to different beam current regimes.
\end{abstract}

\maketitle

\section{INTRODUCTION}
\begin{figure}[tb]
    \centering
    \includegraphics[width=\columnwidth]{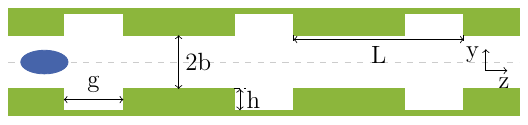}
    \caption{The corrugated plates in cross-section (side view) with the relevant geometric parameters are shown. The electron bunch is indicated in blue, traveling along the $z$ axis.}
    \label{fig:corrugated_pipe}
\end{figure}

In contrast to incoherent synchrotron radiation, coherent synchrotron radiation (CSR) scales not linearly but quadratically with the number of photons. Thus, the emitted photon flux is amplified by several orders of magnitude. The radiation is coherent if the emitted photon wavelength is larger than the radiating structure. To generate CSR pulses in the \si{\tera\hertz} frequency range, the length of the emitting structure has to be reduced to the few-picoseconds time scale.\\
At KARA, this picosecond structure of the electron bunches is realized in optics with reduced momentum compaction factor (low-$\alpha_\text{c}$)~\cite{MullerPAC05, PapashIPAC18}. The resulting high particle density entails the interaction of the electron bunches with their self-emitted CSR. This can lead to longitudinal bunch deformation and dynamic instabilities, like the so-called micro-bunching instability~\cite{Byrd02,Shields12,Brosi21}. It causes longitudinal substructures on the bunches and generates quasi-periodic outbursts of intense \si{\tera\hertz} radiation~\cite{Venturini02, MullerPAC05, RousselPhysRev14}. Understanding and controlling this instability are crucial to meet the requirements of modern electron storage rings in terms of brilliance and peak power by providing a higher electron density. Furthermore, this knowledge allows opening up a new frequency range of intense radiation by extending the coherent emission to higher frequencies.\\
At the Synchrotron SOLEIL storage ring, C.~Evain~\emph{et~al.}~\cite{EvainNature19} use feedback control in single bunch mode at a constant bunch current to study and sustain the micro-bunching structure to produce CSR, which is stable over time. With the steady-state micro-bunching mechanism, X.~Deng~\emph{et al.}~\cite{DengNature21} demonstrated control of the instability and can generate high-power CSR at the Metrology Light Source (MLS) with an active energy-modulation by a laser beam in an undulator.\\
At KARA, a feedback control based on reinforcement learning is used to control the micro-bunching instability~\cite{BoltzIPAC19,ScomparinIBIC22} during in dependence of the beam current. Additionally, an impedance manipulation chamber is currently under development for the KARA storage ring to change the longitudinal impedance of the ring, as well as the longitudinal wake fields. This additional impedance permits influencing the beam dynamics of the passing electrons and, thus, the micro-bunching instability. It is created by a pair of horizontal parallel plates with periodic rectangular corrugations perpendicular to the direction of bunch motion, placed in a straight section of the storage ring. Figure~\ref{fig:corrugated_pipe} shows a schematic drawing of the structure with the parameters of corrugation depth $h$, periodic length $L$, longitudinal gap $g$, and plate distance $2b$. The duty cycle is defined by $L/g$ for these structures. The design of the chamber allows a vertical movement to adjust the impact of the structure and a horizontal movement to chose between different structures and impedances~\cite{MaierIPAC21}. The Frenet-Serret coordinates are defined in the way that $x$ is the transversal horizontal, $y$ is the transversal vertical, and $z$ is the longitudinal direction.
K.~L.~Bane~\emph{et~al.}~\cite{BaneNuclInstr12, BaneNuclInstr17} have already tested a cylindrical corrugated structure in a linear accelerator at Brookhaven's Accelerator Test Facility (ATF). They created narrow-band \si{\tera\hertz} pulses of Smith-Purcell radiation~\cite{SmithPurcell53}. At KARA, we also expect to generate Smith-Purcell radiation, but its investigation is not the primary goal of the project. Its main goal is rather to investigate and understand how the corrugated structure influences the micro-bunching instability and the following CSR generation in conventional sources such as bending magnets at light sources. Such corrugated structures have already been installed and tested as an energy dechirper of linac-based free electron lasers (FELs) at the PAL-XFEL~\cite{EmmaPhysRevDechirper} and SwissFEL~\cite{DijkstalIPAC23Dechirper}. To our knowledge, such a structure has not yet been installed into a storage ring, where the additional passive structure can affect the bunch profile and the emitted CSR frequently with the repetition rate of the revolution frequency. This approach enables an additional knob to control the beam dynamics. Therefore this study does not only discuss the beam control on KARA but also serves as proof of principle for the potential of this method in other machines.\\
Recently, S.~A.~Antipov~\emph{et~al.}~\cite{Antipov22} have simulated the influence of a corrugated structure on the bunch length. They found that the bunch length below the micro-bunching instability threshold can be reduced significantly so that the CSR emission spectrum is extended to higher frequencies. To do this, they considered a structure with a periodic length of the wake corresponding to the bunch length for a KARA low-$\alpha_\text{C}$ operation mode. However, they only took the impedance of the corrugated structure into account but neglected all other beam-coupling impedances that affect the longitudinal bunch distribution. In contrast, in our discussion presented here, we also take the CSR impedance into account, which at KARA is the dominant contribution. Furthermore, we have simulated the influence of different corrugation geometries on the longitudinal beam dynamics and the impact of different impedances versus the machine parameters, namely momentum compaction factor $\alpha_\text{C}$ and acceleration voltage $V_\text{acc}$. These studies contribute to further understanding of the underlying beam physics during the micro-bunching instability by showing the relation between the most effective resonance frequency and the bunch profile.
\section{CORRUGATED PLATE IMPEDANCE}
\begin{figure}[t]
    \centering
    \includegraphics[width=\columnwidth]{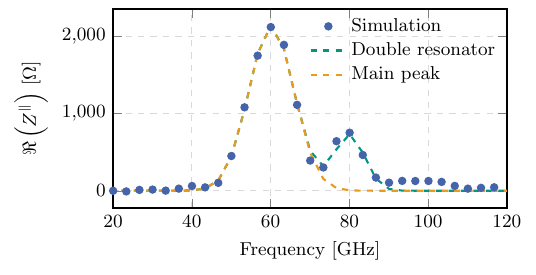}
    \caption{Real part of a simulated corrugated structure impedance. To determine the impedance parameters shunt impedance $Z_0$, quality factor $Q$, and resonance frequency $f_\text{res}$, a superposition of two resonator functions are used for the fit. The dimensions of the corrugations are: half plate distance $b=$\SI{5}{\milli\meter}, corrugation depth $h=$\SI{300}{\micro\meter}, periodic length $L=$\SI{200}{\micro\meter}, duty cycle $L/g=$2, and structure length $s=$\SI{20}{\centi\meter}.}
    \label{fig:impedance_fit}
\end{figure}
\begin{figure}[t]
    \centering
    \includegraphics[width=\columnwidth]{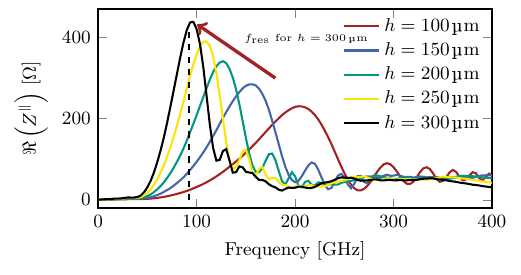}
    \caption{The real part of the longitudinal impedance for different corrugation depths $h$ is simulated with CST Particle Studio. The other corrugation parameters are fixed: half plate distance $b=\SI{5}{\milli\meter}$, periodic length $L=\SI{100}{\micro\meter}$, duty cycle $L/g=4$, and structure length $s=\SI{10}{\centi\meter}$. A deepening of the corrugations causes a reduction of the corrugation resonance frequency, an increase of the shunt impedance, and a narrowing of the resonance peak. The dashed line and the arrow indicate the resonance frequency for $h=\SI{300}{\micro\meter}$ as an example.}
    \label{fig:impedance_hScan}
\end{figure}
\begin{figure}[tb]
    \centering
    \includegraphics[width=\columnwidth]{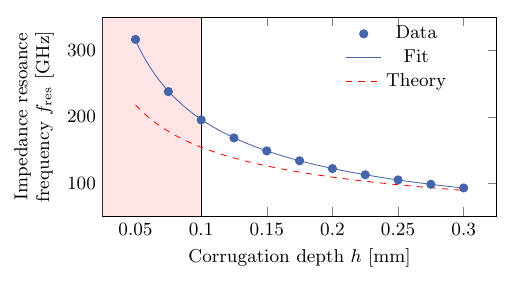}
    \caption{The corrugation resonance frequency versus the corrugation depth $h$ with the same fixed parameters as in Fig.~\ref{fig:impedance_hScan}. The red dashed line is based on the theoretical prediction for the parallel plate geometry. Note that the validity condition of this equation is not met in the red shaded area.}
    \label{fig:impedance_hScan_fresFit}
\end{figure}

The theory of the longitudinal impedance $Z^\parallel$ of a cylindrical corrugated pipe is described by  Ng and Bane in~\cite{ChaoHandbook} with the validity range $L\lesssim h\ll b$, as
\begin{equation}
\frac{Z^\parallel}{L}=\frac{Z_\text{vac}}{\pi b^2}\left[\pi k_{\text{res}}\delta\left(k^2-k_{\text{res}}^2\right)+i \text{P.V.}\left(\frac{k}{k^2-k_{\text{res}}^2}\right)\right]\label{eq:impedance_corrugated_pipe}
\end{equation}
with the resonance wave number $k_{\text{res}}=\sqrt{\frac{2L}{bgh}}$, the wave number $k=\frac{\omega}{c}$~\cite{BaneTheory}, the vacuum impedance $Z_\text{vac}\approx\SI{377}{\ohm}$, the Dirac $\delta$-distribution, and the principal value P.V.(x)~\cite{Vladimirov71}. The relation between the resonance wave number and resonance frequency is given by $f_\text{res}=\frac{k_\text{res}}{2\pi c}$.\\
For the KARA storage ring, however, a structure consisting of two parallel plates is being developed instead of the cylindrical geometry. For this structure, the resonance frequency is known as $f_\text{res}=\frac{1}{2\pi c}\sqrt{\frac{L}{bgh}}$~\cite{Bane03} and thus smaller by the factor $\sqrt{2}$ than for the cylindrical geometry. Since the corrugation parameter range investigated is not limited to the validity range of the theoretical formula, the resonance frequency needs to be determined by simulations. Furthermore, the formula contains no information about the shunt impedance $Z_0$ and the finite width of the resonance peak. For the simulations of the corrugated structure impedance, the Wakefield Solver of CST Particle Studio~\cite{CST} is used. The simulated structure consists of two identical and mirrored corrugated plates out of electrically conductive material (here: stainless steel) and a vacuum in between.  A more detailed description of the simulation settings is given in Ref.~\cite{MaierIPAC21}. As discussed in Ref.~\cite{MaierIBIC22}, a width of the corrugated strips of at least $x_0=\SI{20}{\milli\meter}$ is necessary to avoid the edge effect, which comes from the horizontal boundary of the corrugated plates. For the simulations shown here, the width is set to $x_0=\SI{24}{\milli\meter}$.\\
The real part of the simulated impedance $\Re\left(Z^\parallel\right)$ of the corrugated plates is shown in Figure~\ref{fig:impedance_fit}. It can be seen that the impedance can be described with a broadband resonator model~\cite{ChaoBook}
\begin{equation}
    Z(f)=\frac{Z_0}{1+iQ\left(f/f_\text{res}-f_\text{res}/f\right)}\label{eq:resonator_model}
\end{equation}
which is defined by the shunt impedance $Z_0$, the quality factor $Q$, and the corrugation resonance frequency $f_\text{res}$. To improve the description of the main peak and to take the side-peak in the impedance in Figure~\ref{fig:impedance_fit} into account, the impedance is fitted by a superposition of 2 resonators.\\
In Figure~\ref{fig:impedance_hScan} the real part of the simulated impedance $\Re\left(Z^\parallel\right)$ is shown as an example for different corrugation depths $h$ with a fixed half plate distance $b=\SI{5}{\milli\meter}$, periodic length $L=\SI{100}{\micro\meter}$, the duty cycle $L/g=4$, and the structure length $s=\SI{10}{\centi\meter}$. It can be seen that a single parameter of the corrugation geometry affects all impedance parameters: a deepening of the corrugations leads not only to a decrease of $f_\text{res}$ but also to an increase of the shunt impedance and a narrowing of the peak.\\
The corrugation resonance frequency of the main peak - determined by the 2-resonators fit - versus the corrugation depth is shown in Figure~\ref{fig:impedance_hScan_fresFit}. The white area indicates the validity range of Eq.~\ref{eq:impedance_corrugated_pipe} $\left(L\lesssim h\right)$. Besides the theoretical dependency for the parallel plate geometry, a fit with the behavior $1/h^\gamma$ is also shown. For the corrugation settings shown here, the slope is $\gamma=\SI{0.684\pm 0.002}{}$, which is comparable but significantly larger than theoretical prediction ($\gamma_\text{theo}=0.5$). From Figures~\ref{fig:impedance_hScan} and \ref{fig:impedance_hScan_fresFit} it can be seen that with increasing corrugation depth h and resulting larger shunt impedance the deviations from a broad-band resonator model decrease and the resonance frequency converges to the theoretical prediction.\\
The available space at the KARA storage ring allows a maximal length of $s=\SI{20}{\centi\meter}$ for the corrugated plate. This results in an achievable shunt impedance of $Z_0=\SI{1}{\kilo\ohm}$, which has been used throughout the beam dynamics studies~\cite{MaierPhD}. The deviation from the resonator model is in this case negligible. From the systematic impedance simulations in dependence of all parameters of the corrugation \cite{MaierIPAC21}, the geometry of the corrugated structure that creates a certain impedance can be calculated. 
\section{BEAM DYNAMICS SIMULATION}
\begin{table}[tb]
   \centering
   \caption{KARA settings used in Inovesa simulations}
   \begin{ruledtabular}
   \begin{tabular}{lcc}
     \textbf{Parameter}& \textbf{Value}  & \textbf{Unit}          \\\hline
       KARA circumference & 110.4 & \si{\meter}\\
       Bending radius $\rho$ & 5.559 &\si{\meter}\\
       Vacuum chamber height $h_\text{c}$ & 32 &\si{\milli\meter}\\
       Electron beam energy $E$  & 1.3&\si{\giga\electronvolt} \\
       RF frequency $f_\text{rf}$ & 499.705&\si{\mega\hertz}  \\
       Revolution frequency $f_\text{rev}$ & 2.7157&\si{\mega\hertz}  \\
       Harmonic number $h$&184 & \\  
       Synchrotron frequency $f_\text{s}$ & 8.87 to 15.90 &\si{\kilo\hertz} \\
       RF voltage $V_\text{acc}$ & 800 to 1400 &\si{\kilo\volt} \\
       Momentum  compaction factor $\alpha_\text{c}$ & 6 to \SI{13e-4}{}&\\
       Radiated energy/particle/revolution $U_0$ & 45.5 & \si{\kilo\electronvolt}\\
   \end{tabular}
   \label{tab:kara_parameters}
   \end{ruledtabular}
\end{table}
The Vlasov-Fokker-Planck solver Inovesa~\cite{InovesaCode} has been used for the longitudinal beam dynamics simulations. P.~Sch\"onfeldt~\emph{et~al.}~\cite{Inovesa} have shown that Inovesa describes the properties of the micro-bunching instability properly. Thereby, it has been shown that the beam dynamics of the micro-bunching in the KARA storage ring can be described in excellent agreement with the experimental results by only considering the shielded CSR impedance with the parallel plate model. Thus, the other ring impedances are negligible in this context. Therefore, a change in the beam dynamics due to an additional corrugated plate impedance also shows that the effect of this impedance on the micro-bunching instability is significantly stronger than that of other geometric impedances at KARA. The relevant machine settings for the low-$\alpha_\text{c}$ mode at KARA are given in Table \ref{tab:kara_parameters} and are also used for the Inovesa simulations. For these simulations, two of three parameters related to the longitudinal beam dynamics must be given: namely, the synchrotron frequency $f_\text{s}$, the momentum compaction factor $\alpha_\text{c}$, and the acceleration voltage $V_\text{acc}$. They are related via the Equation~\cite{Wiedemann15}:
\begin{equation}
    f_\text{s}=f_\text{rev}\sqrt{\frac{h\alpha_\text{c}}{2\pi E}\sqrt{e^2V_\text{acc}^2-U_0^2}}\label{eq:SychrotronFrequency}
\end{equation}
 In the case of a scan over different impedances with the same machine settings and beam current, the phase space distribution of the same settings without an additional impedance is used as the initial distribution. In contrast, the initial distribution for a current scan is taken from the final state distribution of the previous and higher current step to model the current decay at KARA. For the results in this paper, the simulated longitudinal bunch profile and the CSR power over time have been considered. Since the micro-bunching instability around the threshold current is to be examined, the dynamics are in a non-equilibrium state. However, the simulation needs some preparation steps to converge, so the first 150 synchrotron periods \ensuremath{T_\text{S}} are ignored as settling time, which is the time range for the convergence. The following \SI{350}{\ensuremath{T_\text{S}}} are investigated for the analyses. \\
The dominating CSR parallel plate impedance describes the overall impedance of the KARA storage ring \cite{Murphy97} according to the preceding studies for describing the micro-bunching instability at KARA \cite{Inovesa, BrosiPhysRev16}.\\
In the following studies, an additional impedance according to the corrugated plate model has been added to the CSR impedance for the beam dynamics simulations. Thereby, a scan over an impedance parameter is more useful than one over the corrugation geometry to understand the mechanism which drives the micro-bunching instability. The geometrical parameters of the corrugated structure can be defined by the impedance parameters established in the simulations. So the added impedance is not based on a defined corrugation geometry but is given by Eq.~\ref{eq:resonator_model} and is defined by $Z_0$, $Q$, and $f_\text{res}$.
A systematic study of the corrugation structure impedance~\cite{MaierIPAC21} has shown that $Q=\SI{3}{}$ and $Z_0=\SI{1}{\kilo\ohm}$ are suitable for different corrugation geometries for a structure length of $s=\SI{20}{\centi\meter}$, which is the maximal space available for the corrugated plates in the KARA storage ring. Therefore, unless otherwise noted, these parameters are set to $Q=\SI{3}{}$ and $Z_0=\SI{1}{\kilo\ohm}$. In the frequency range between $\SI{50}{\giga\hertz}$ and $\SI{200}{\giga\hertz}$, which is used for the impedance resonance frequency in this paper, the CSR impedance for KARA is at least $Z_\text{csr}=\SI{6}{\kilo\ohm}$. 
Consequently, the impedance of the corrugated structure is only a perturbation in the order of up to \SI{10}{\percent}. The side peak (see Figure~\ref{fig:impedance_fit}) is negligible compared to the CSR impedance. Therefore, only a single resonator impedance is added for the beam dynamics simulations.
\section{CORRUGATION RESONANCE FREQUENCY SCAN}\label{sec:ImpedanceScan}
\begin{figure*}[tb]
    \centering
    \includegraphics*[width=\textwidth]{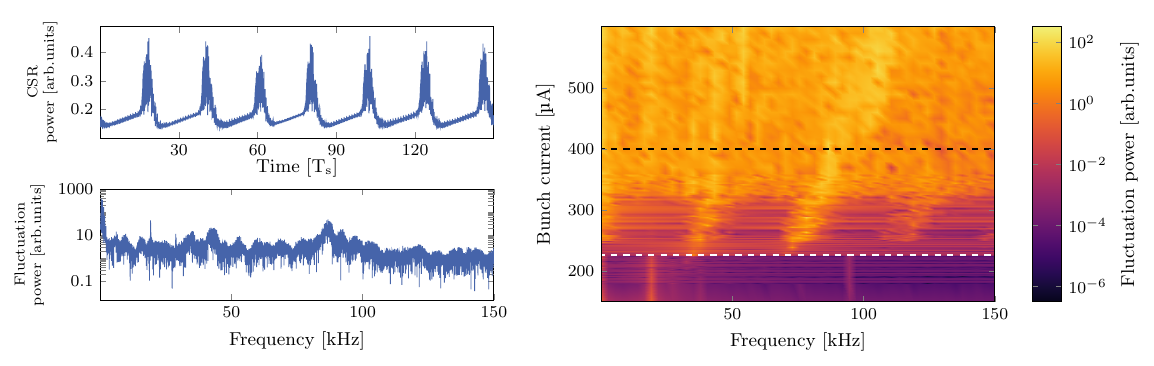}
    \caption{Top left: Simulated emitted CSR power as a function of time in terms of the synchrotron period $T_\text{S}$ at a beam current of $I=\SI{400}{\micro\ampere}$ in the bursting regime (without additional corrugation plate impedance). Bottom left: The Fourier transform of the CSR power indicates the dominant micro-bunching frequencies. Right: The spectrogram shows the color-coded fluctuation power for different bunch currents as a function of the frequency. The black dashed line indicates the bunch current of $I=\SI{400}{\micro\ampere}$, for which the plots on the left are shown, and the white dashed line indicates the threshold current $I_\text{thr}=\SI{226}{\micro\ampere}$. The constant machine settings for this simulation are $f_\text{s}=\SI{9.44}{\kilo\hertz}$ and $V_\text{acc}=\SI{1.048}{\mega\volt}$.}
    \label{fig:intensity_to_spectrogram}
\end{figure*}
\begin{figure}[b]
    \centering
    \includegraphics[width=\columnwidth]{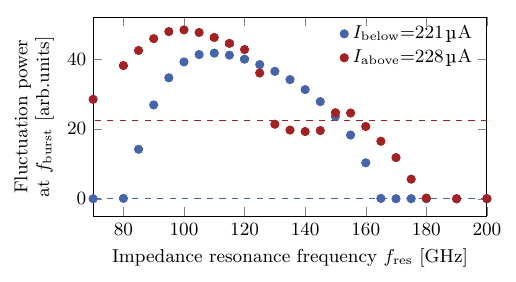}
    \caption{Fluctuation power at the dominant frequency as a function of the corrugation resonance frequency of an additional impedance for a bunch current slightly below (blue) and above (red) the unperturbed bursting threshold current. The dashed lines indicate the fluctuation power without an additional impedance. Constant parameters are $Z_0=\SI{1}{\kilo\ohm}$, $Q=3$, $f_\text{s}=\SI{9.44}{\kilo\hertz}$, and $V_\text{acc}=\SI{1.048}{\mega\volt}$.}
    \label{fig:ImpedanceScan}
\end{figure}

For a systematic investigation of the corrugated structure impedance's impact on the longitudinal beam dynamics, the machine settings $f_\text{s}=\SI{9.44}{\kilo\hertz}$ and $V_\text{acc}=\SI{1.048}{\mega\volt}$ of a dedicated and well-examined KARA setting with the zero-current bunch length $\sigma_0=\SI{4.1}{\pico\second}$ are used. This study inspects the bunch current range around the unperturbed bursting threshold current $I_\text{thr}$. Without an additional impedance, the micro-bunching instability develops above  $I_\text{thr}$. Thus, the first fluctuations of the emitted CSR occur directly above the threshold. Therefore, the standard deviation (STD) of the CSR power as a function of the bunch current shows a sharp change at the bursting threshold current~\cite{BrosiIPAC17}, which is used to determine $I_\text{thr}=\SI{226}{\micro\ampere}$.\\

The top left part of Figure~\ref{fig:intensity_to_spectrogram} shows the emitted CSR power versus the time in the bursting regime. In the Inovesa simulations, the CSR power is calculated from the spectrum of the longitudinal bunch profile $\tilde{\varrho}$ and the real part of the impedance: $P_\text{CSR}(t)\propto\int \Re\left({Z_k}\right)\times\left|\tilde{\varrho}(k,t)\right|^2\text{d}k$, which is derived from the charge distribution in the longitudinal phase space~\cite{Inovesa}. The Fourier transform of the simulated CSR power, which is called fluctuation power, reveals the fluctuation frequencies of the CSR power and therefore points out the dominant micro-bunching frequencies (bottom left part of Figure~\ref{fig:intensity_to_spectrogram}). This corresponds to one line (dashed line) in the spectrogram on the right side of the Figure, which shows the fluctuation power as a function of the beam current. The finger-like structure in the frequency range from \SIrange{30}{40}{\kilo\hertz} originates from the creation of the substructures above the threshold $I_\text{thr}$~\cite{BrosiPhysRev16} and causes a sinusoidal fluctuation in the CSR power. This frequency is determined by the rotation of the substructures in the longitudinal phase space. The other dominant frequency below \SI{2}{\kilo\hertz} indicates the slow bursting frequency, corresponding to the repetition rate of the temporal generation and smearing of the substructures in the longitudinal bunch profile. These temporal outbursts are self-induced by the micro-bunching instability at bunch currents above $I_\text{slow}$ and considerably above the threshold current $I_\text{thr}$~\cite{BrosiPhD}.\\
For studying how the micro-bunching instability threshold is affected by an additional corrugated plate impedance, two currents which are slightly above ($I_\text{above}=\SI{228}{\micro\ampere}$) and slightly below ($I_\text{below}=\SI{221}{\micro\ampere}$) the unperturbed threshold have been chosen. The temporal development of the emitted CSR power with varying $f_\text{res}$ of the additional impedance has been simulated for these two beam current conditions. Figure~\ref{fig:ImpedanceScan} shows the fluctuation CSR power at the bursting frequency depending on $f_\text{res}$ of the additional impedance for the two bunch currents. In this context, the bursting frequency $f_\text{burst}$ is determined as the frequency where the fluctuating CSR power is maximal.\\
Without an additional impedance, no significant CSR power is emitted below the unperturbed bursting threshold (blue dashed line in Figure~\ref{fig:ImpedanceScan}). However, with a specific corrugated plate impedance, the radiation of intense CSR power can be enhanced significantly so that the emitted intensity is comparable to or even higher than the maximal power directly above the threshold current (red dashed line in Figure~\ref{fig:ImpedanceScan}) without a corrugated plate impedance. This increase of CSR power is only reached in the case of the corrugation resonance frequency in the range from \SIrange{85}{160}{\giga\hertz}. These impedances reduce the bursting threshold below $I_\text{below}$ and cause the micro-bunching at even lower bunch currents.\\
For the bunch current $I_\text{above}$, the maximal fluctuation power is enhanced for $f_\text{res}\leq\SI{130}{\giga\hertz}$. The maximal amplification is reached with $f_\text{res}=\SI{110}{\giga\hertz}$ and more than doubled in comparison with the pure CSR impedance. In contrast, a corrugation resonance frequency above \SI{160}{\giga\hertz} leads to a reduction and above \SI{180}{\giga\hertz} even to a complete suppression of the emitted CSR power in the \si{\tera\hertz} range.

\section{THRESHOLD CURRENT \& BURSTING FREQUENCY}
\begin{figure*}[bt]
    \centering
    \includegraphics*[width=\textwidth]{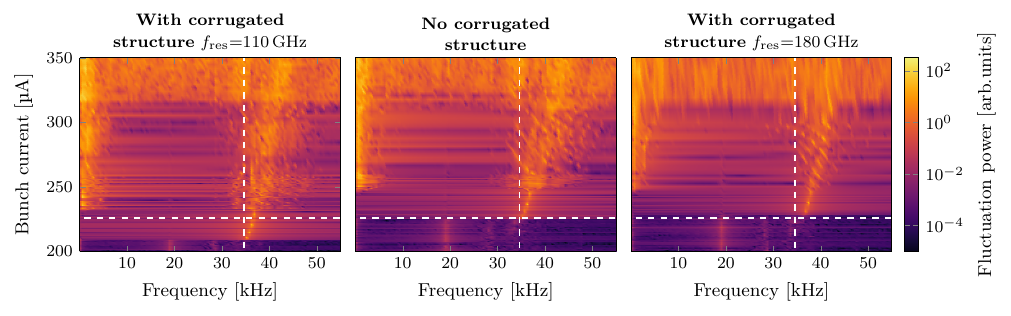}
    \caption{Color-coded Fourier transform of the emitted CSR power for different bunch currents with and without additional corrugated plate impedances for two different corrugation resonance frequencies $f_\text{res}$. The dashed white lines mark the threshold current of the micro-bunching and the dominant bursting frequency $f_\text{burst}$ at the threshold for the case with no additional impedance (middle) as a reference. The shunt impedance and the quality factor are fixed at $Z_0=\SI{1}{\kilo\ohm}$ and $Q=3$, respectively.}
    \label{fig:spectrograms}
\end{figure*}
\begin{figure}[bt]
    \centering
    \includegraphics[width=\columnwidth]{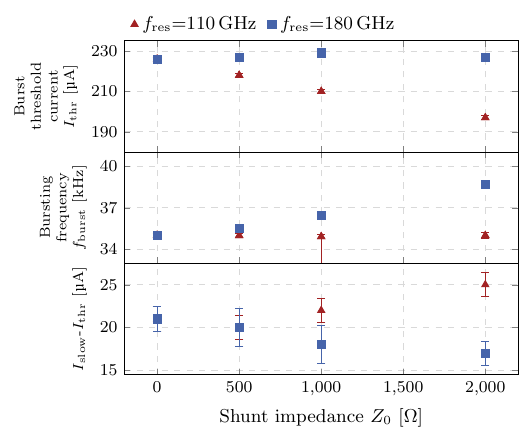}
    \caption{Bursting threshold parameters versus the shunt impedance of the additional impedance. An impedance with the corrugation resonance frequency $f_\text{res}=\SI{110}{\giga\hertz}$ (red) reduces the threshold current, whereas $f_\text{res}=\SI{180}{\giga\hertz}$ (blue) changes the bursting frequency. Since the impact of $f_\text{res}=\SI{110}{\giga\hertz}$ on the bursting threshold current $I_\text{thr}$ is stronger than on the slow bursting threshold $I_\text{slow}$, the current range between the two regimes can be increased (bottom). The machine settings are the same as in Figure~\ref{fig:ImpedanceScan}.}
    \label{fig:FingerParamter_Z0scan}
\end{figure}

In the previous section, it has been shown that the threshold of micro-bunching can be changed depending on $f_\text{res}$ of an additional corrugated plate impedance. Two dedicated additional impedances are chosen for a more elaborated investigation of the impedance impact on the longitudinal beam dynamics. The emitted CSR power in dependency on the bunch current has been simulated for those impedances. The first selected impedance is the one for which the fluctuation of the CSR emission is most enhanced ($f_\text{res}=\SI{110}{\giga\hertz}$). The second one is given as the higher edge ($f_\text{res}=\SI{180}{\giga\hertz}$), above which the fluctuation of the CSR ceases slightly above the unperturbed threshold current (see the red circles in Figure~\ref{fig:ImpedanceScan}). For those impedances and the reference with only the CSR impedance, the Fourier transform of the simulated emitted CSR power as a function of the bunch current is shown in Figure~\ref{fig:spectrograms} as spectrograms.\\
An additional impedance with $Z_0=\SI{1}{\kilo\ohm}$ significantly changes $I_\text{thr}$ and $f_\text{burst}$ without substantially changing the overall shape of the structures in spectrograms and the bursting behavior of the substructures. The simulation results show that the impedance of an additional corrugated structure with $f_\text{res}=\SI{110}{\giga\hertz}$ reduces the threshold current by \SI{16}{\micro\ampere} (\SI{7.1}{\percent}). In contrast, the dominant bursting frequency at the threshold current is unchanged. \\
In opposite, an impedance with $f_\text{res}=\SI{180}{\giga\hertz}$ not only increases the threshold slightly by \SI{3}{\micro\ampere} (\SI{1.3}{\percent}) but also shifts $f_\text{burst}$ by +\SI{1.9}{\kilo\hertz} to a higher repetition rate. Even though these changes are small, they are measurable with the diagnostics available at KARA. It need to be mentioned, that this impact can be reached by a very compact structure, that only covers \SI{0.18}{\percent} of the KARA circumference.\\
Figure~\ref{fig:FingerParamter_Z0scan} illustrates how the shunt impedance and, therefore, the share of the corrugated structure in the total KARA impedance affects the properties of the micro-bunching bursting regime for the two $f_\text{res}$ settings. The top two plots show that the two bursting parameters $f_\text{burst}$ and $I_\text{thr}$ scale nearly linearly with $Z_0$ for one of the two different corrugation resonance frequencies (\SI{110}{\giga\hertz}: red, \SI{180}{\giga\hertz}: blue). In contrast, the other impedance does not affect the same parameter significantly~\cite{MaierIBIC22}. The third plot reveals that the threshold currents of the two bursting regimes ($I_\text{thr}$ and $I_\text{slow}$) are not changed with the identical strength by the additional impedances. It follows that for the impedance with $f_\text{res}=\SI{110}{\giga\hertz}$ the gap between the two thresholds can be increased significantly. In combination with the results, that the shunt impedance scales linearly with the length of the corrugated structure~\cite{MaierPhD}, it follows that a structure with, e.g. $s=\SI{1}{\meter}$, which corresponds to \SI{0.9}{\percent} of the KARA circumference could cause a reduction of the bursting threshold by \SI{30}{\percent}.\\
By C.~Evain \emph{et al.}~\cite{Evain12}, this is called the stable regime since the fingers in the longitudinal phase space only rotate and do not change their intensity periodically. So enhanced CSR in the \si{\tera\hertz} range is emitted without the temporal outbursts. Consequently, such an impedance can be used to increase the current range in which intense and stable CSR is created in the low-$\alpha_\text{c}$ operation mode at KARA.
\section{MACHINE SETTING SCAN}\label{sec:machineSettings}
\begin{figure*}[tb]
    \centering
    \includegraphics[width=\textwidth]{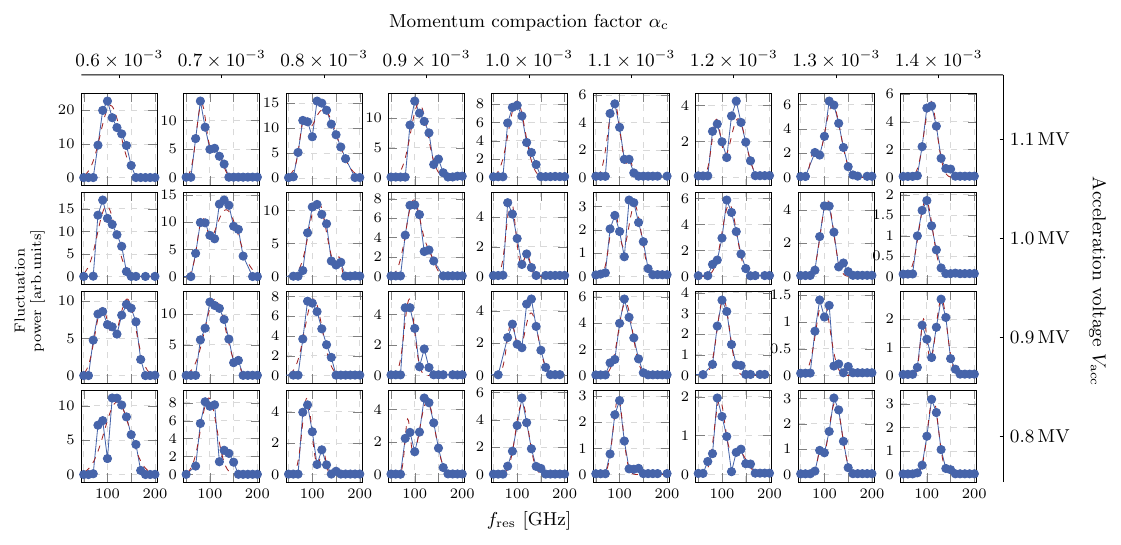}
    \caption{The fluctuation power at the bursting frequency $f_\text{burst}$ as a function of the corrugation resonance frequency of an additional impedance is shown for nine $\alpha_\text{c}$ and four $V_\text{acc}$ settings. The current is below the bursting regime with \SI{98.5}{\percent} of the respective unperturbed bursting threshold. For the fit (dashed red line), a 2-peak Gaussian is used. Note that the second peak is only visible for some machine settings.}
    \label{fig:2DImpedanceScan}
\end{figure*}
\begin{figure}[tb]
    \centering
    \includegraphics[width=\columnwidth]{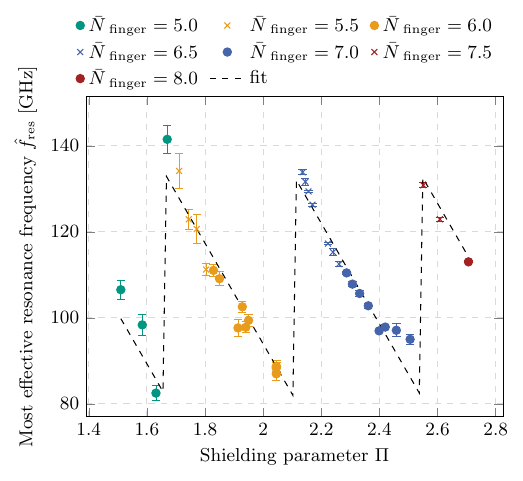}
    \caption{Most effective corrugation resonance frequency $\hat{f}_\text{res}$ to reduce the bursting threshold as a function of the shielding parameter $\Pi$ (corresponding to the bunch length). The colors and shapes denote the number of fingers in the longitudinal phase space (see Fig.~\ref{fig:phaseSpace}). The fit is based on a saw-tooth function. The green outlier is caused by the automated counting of substructures since the intensity of one of the fingers is not high enough to reach the threshold for the counting.}
    \label{fig:SettingScan_bunchLength}
\end{figure}
\begin{figure}[tb]
    \centering
    \includegraphics[width=\columnwidth]{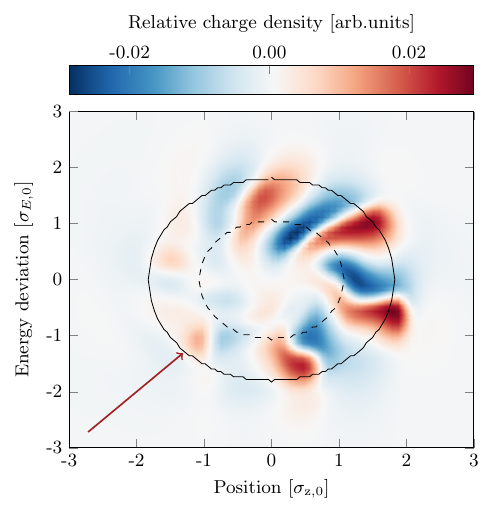}
    \caption{Longitudinal phase space with non-integer number of fingers $\bar{N}_\text{finger}=5.5$. Subtracting the mean temporal distribution of the phase space density emphasizes the fingers. The fingers develop from the outside (solid line) to the inside (dashed line). In the case shown, the additional finger in the left bottom corner (red arrow) is only present in the outer part of the phase space, leading here to a non-integer number of fingers. The machine settings are $V_\text{acc}$=\SI{0.9}{\mega\volt} and $\alpha_\text{c}$=\SI{7e-4}{}.}
    \label{fig:phaseSpace}
\end{figure}
\begin{figure}[tb]
    \centering
    \includegraphics[width=\columnwidth]{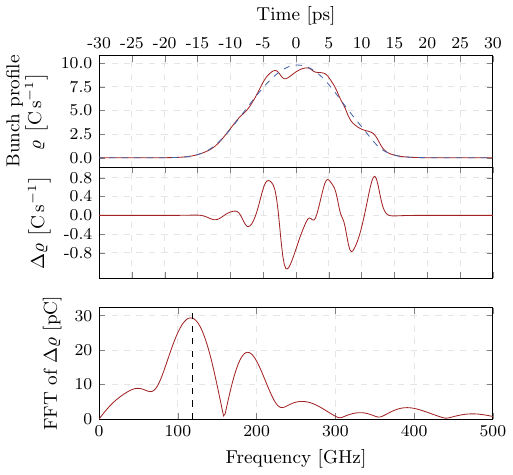}
    \caption{Bunch profile (red) at the simulation time step with the largest CSR emission power and the temporal averaged bunch profile (blue dashed) as a reference (top). The substructures in the longitudinal bunch profile become more visible after subtracting the temporal average profile (middle). The Fourier transform of this relative bunch profile $\Delta\varrho$ indicates the dominant frequency of its spectrum, which is in good agreement with $\hat{f}_\text{res}$ (dashed black line). The machine settings are $V_\text{acc}$=\SI{0.8}{\mega\volt} and $\alpha_\text{c}$=\SI{6e-4}{}.}
    \label{fig:BunchProfileSpectrum}
\end{figure}

For a further and methodical understanding of the mechanisms driving the micro-bunching instability, the impact of an additional impedance on the threshold current has been studied for a systematic scan of machine parameters $V_\text{acc}$ and $\alpha_\text{c}$. Since the threshold current depends directly on both machine parameters, the bunch current in the simulations has been adjusted accordingly. It has been chosen as \SI{98.5}{\percent} of the respectively unperturbed bursting threshold current to see the impact of the corrugated plate impedance on the behavior of the micro-bunching instability around the threshold.\\
Figure~\ref{fig:2DImpedanceScan} shows the maximal fluctuation power versus $f_\text{res}$ for all the simulated settings with a matrix form. If the main peak is below $\hat{f}_\text{res}=\SI{100}{\giga\hertz}$ a second peak occurs at about \SI{170}{\giga\hertz}. By increasing $\alpha_\text{c}$, the peak at the lower $f_\text{res}$ decreases until it becomes insignificant, and the other one increases in fluctuation power until it becomes dominant at about $\hat{f}_\text{res}=\SI{135}{\giga\hertz}$. Moreover, a double-peak Gaussian can describe all frequency scans, where the second peak only occurs for some machine setting configurations. From the curve fitting with this function indicated by the red dashed line, the most effective corrugation resonance frequency $\hat{f}_\text{res}$ of the additional impedance for reducing the $I_\text{thr}$ can be determined.\\
It can be seen that the threshold current is only affected significantly by $f_\text{res}$ in the range between \SIrange[range-phrase=\text{ and }]{50}{170}{\giga\hertz}, regardless of the settings. Since the investigated currents of the different machine settings are not comparable and are in the range from \SIrange{262}{937}{\micro\ampere}, the comparison of the amplitude of the fluctuation power at $f_\text{burst}$ is not meaningful. In the matrix form, it appears that the $\hat{f}_\text{res}$ and the curve shapes are nearly identical along the rising diagonal on which the zero-current bunch length is nearly constant.\\
However, the knowledge about the impact of the machine settings on $\hat{f}_\text{res}$ is important for operating a real machine with a certain corrugated structure. Additionally, the influence of the beam parameters is crucial for understanding and insight into the driving processes of the micro-bunching instability. Therefore, the machine settings are reduced to the dimensionless and accelerator-independent shielding parameter $\Pi$ \cite{Murphy97}
\begin{equation}
    \Pi=\frac{\rho^\frac{1}{2}}{2 h_{1/2}^\frac{3}{2}}\sigma_\text{z,0}\label{eq:shielding_parameter}
\end{equation}
with the bending radius $\rho$, the half height of the vacuum chamber $h_{1/2}=\frac{1}{2} h_\text{c}$, and which is proportional to the zero-current bunch length $\sigma_\text{z,0}$.\\
In Figure \ref{fig:SettingScan_bunchLength}, the most effective corrugation resonance frequency is shown as a function of the shielding parameter. The color- and shape code indicate the number of the fingers $\bar{N}_\text{finger}$ in the longitudinal phase space density and bunch profile caused by the micro-bunching instability \cite{MaierIPAC23}.
For counting the fingers, the charge density along two circles (dashed and solid circle in Figure~\ref{fig:phaseSpace}) in the longitudinal phase space is analyzed. For better recognition of the fingers, the time-averaged phase space is subtracted. The two circles - representing the inner and outer parts of the phase space - are defined as the first and last circles whose maximal charge density is larger than 0.3 of the maximal charge density in the total phase space. Along these lines, the number of sign changes are counted to determine the number of fingers. The half-integer number of fingers implies a different number of fingers in the inner and outer parts of the phase space, shown as an example in Figure~\ref{fig:phaseSpace}. In the left bottom corner, it can be seen that in the outer part (solid line), a finger is already formed, whereas, in the inner part (dashed line), this finger can not be seen. For the determination of $\bar{N}_\text{finger}$ the fingers in the phase space at the time of maximum CSR emission during the micro-bunching outburst and at a current \SI{100}{\micro\ampere} above $I_\text{thr}$ without an additional impedance are counted. The counting has been cross-checked with the ratio $\frac{f_\text{burst}}{f_\text{s}}$, whose connection with the number of fingers has been verified by the formula $\bar{N}_\text{finger}=\frac{f_\text{thr}}{f_\text{s}}+1$ \cite{BrosiPhD} for the investigated area of $\Pi$ at KARA.\\
It can be seen that $\hat{f}_\text{res}$ decreases nearly linearly while the mark-color is constant, which corresponds to a constant number of fingers in the outer part of the phase space. From the fit with a sawtooth function follows, that the decrease of $\hat{f}_\text{res}$ goes down to $f_\text{res,min}=\SI{82.9\pm 3.2}{\giga\hertz}$. However, as soon as the bunch is long enough so that an additional finger can be formed, the corrugation resonance frequency of the most effective additional impedance jumps to the maximal value $f_\text{res,max}=\SI{132.8\pm 1.4}{\giga\hertz}$. In this transition range, the fluctuation power as a function of the corrugation resonance frequency has two peaks, whose amplitudes are nearly identical. T.~Boltz \cite{BoltzPhD} has found a similar sawtooth behavior for the number of fingers as a function of the shielding due to the vacuum chamber height.\\
Beyond that, the fitting with the saw-tooth function
\begin{equation}
    f(x)=-a\text{mod}\left(x-b,x\right)+d
\end{equation}
allows the determination of the dimensions of the fingers with the Euclidean division mod(x,y)=x mod y.
The minimal necessary longitudinal space to form an additional finger is given by the periodicity $\Delta\Pi=\SI{0.443\pm 0.011}{}$ of the saw-tooth or accordingly for KARA by $\Delta\sigma_0=\SI{1.26\pm 0.031}{\pico\second}$. Furthermore, the periodicity of $\hat{f}_\text{res}$ gives the possibility to measure the size of the micro bunches in the KARA storage ring by scanning the machine settings with a fixed installed corrugated structure and measuring the intensity of the emitted synchrotron radiation. The green outlier is caused by the automated counting of substructures since the intensity of one of the fingers is not high enough to reach the threshold for the counting.\\
All this indicates that the additional external impedance directly acts on the micro-bunching substructures. In Figure~\ref{fig:BunchProfileSpectrum}, the longitudinal bunch profile $\varrho$ at a bunch current in the bursting regime is shown (top). It shows the substructures at the simulation time step with the largest CSR\color{red} \color{black} emission power. To make these substructures in the bunch profile more visible, the deviation from the temporal averaged bunch profile is displayed as well (middle). The fast Fourier transform (FFT) of this relative bunch profile $\Delta\varrho$ reveals the dominant frequency of its spectrum (bottom of the Figure). It can be seen that this most dominant frequency is in good agreement with the most effective resonance frequency $\hat{f}_\text{res}$ to manipulate the bursting threshold. This relation is independent of the machine settings and, therefore, is adaptable to other synchrotron light sources to enhance THz radiation through micro-bunching.


\section{CORRUGATED STRUCTURE}
It is planned to install the plates with corrugated structures inside the vacuum pipe of the KARA storage ring. Therefore, they have to be conductive, non-magnetic, and ultra-high vacuum compatible. The vertical movement of the plates enables the adjustment and control of the shunt impedance \cite{MaierIPAC21} as a tuning knob during the machine operation and the deactivation of the additional impedance. For the available space and the width of the KARA beam pipe, three strips with different structures can be placed next to each other with the required minimal width $x_0=\SI{20}{\milli\meter}$ for each strip. This makes it possible to exchange the corrugated structures and their impedance by a horizontal movement without breaking the vacuum. Two of these strips will have different corrugated structures, but one will be a plane plate without corrugations as a reference. The reference strip will be used to distinguish the influence of the corrugations from that of the change in the beam pipe aperture and the additional parallel plate impedance effect. \\
The dimensions of the corrugations have been chosen such that the one corrugation resonance frequency is $f_\text{res,1}=\SI{110}{\giga\hertz}$, which corresponds to $\hat{f}_\text{res}$ of the well-examined KARA machine setting, which has been discussed in Section \ref{sec:ImpedanceScan}. The second strip will have corrugations that produce an impedance with $f_\text{res}=\SI{180}{\giga\hertz}$, which allows the study of the impedance impact on the fluctuation frequency. The optimized parameters of the corrugation to achieve a maximized shunt impedance are given in Table \ref{tab:corrugation_parameters_110_180} for both impedances.\\
The ongoing production of prototypes has pointed out that manufacturing methods based on material removal cause a taper angle at the corrugation sidewalls. Simulation results show the importance of reducing this angle to below \SI{5}{\degree}. Otherwise, $Z_0$ and $f_\text{res}$ will be outside the acceptable ranges. Moreover, the structures will be electrically characterized by measuring the transmission and reflection of electromagnetic waves with proper frequency ranges.

\begin{table}[tb]
   \centering
   \caption{Corrugation and plate parameters for the two chosen structures that create impedances with $f_\text{res,1}$=\SI{110}{\giga\hertz} and $f_\text{res,2}$=\SI{180}{\giga\hertz}, respectively.}
   \begin{ruledtabular}
   \begin{tabular}{lccc}
     \textbf{Parameter}     &  \textbf{Variable}    &   \textbf{$f_\text{res,1}$=\SI{110}{\giga\hertz}}     &   \textbf{$f_\text{res,2}$=\SI{180}{\giga\hertz}}    \\\hline
      Half plate distance   &   b                   &   \SI{5}{\milli\meter}                   &   \SI{5}{\milli\meter}                   \\ 
     Periodic length        &   L                   &   \SI{85}{\micro\meter}                 &   \SI{200}{\micro\meter}                    \\ 
     Corrugation depth      &   h                   &   \SI{90}{\micro\meter}                 &   \SI{55}{\micro\meter}                     \\ 
     Corrugation width      &   g                   &   \SI{59}{\micro\meter}               &   \SI{100}{\micro\meter}                    \\ 
   \end{tabular}
   \label{tab:corrugation_parameters_110_180}
\end{ruledtabular}
\end{table}
\section{SUMMARY}
The installation of a versatile impedance manipulation chamber into the KARA storage ring is planned. The purpose of controlling the machine impedance is to manipulate the longitudinal beam dynamics and thereby study the micro-bunching instability. A pair of parallel, electrically conductive plates with periodic rectangular corrugations will create the additional impedance. The example of the KARA storage ring is used to get a further understanding of the physics of the micro-bunching instability. Furthermore, these studies can be considered as a proof of principle for the manipulation and control of the beam dynamics with a compact and passive structure.\\
The simulations with the Vlasov-Fokker-Planck solver Inovesa show that the corrugation resonance frequency of the additional impedance is the most crucial parameter to manipulate the longitudinal beam dynamics. By choosing the right corrugation resonance frequency, it is possible to manipulate the threshold current of the micro-bunching instability or change its dominant fluctuation frequency. \\
Within the machine setting range of the low-$\alpha_\text{c}$ operation at KARA, the instability threshold is reduced most effectively for a corrugation resonance frequency between \SIrange[range-phrase=\text{ and }]{82.9}{132.8}{\giga\hertz}, so that intense \si{\tera\hertz} radiation is emitted at even lower bunch currents. It could be shown that the corrugation resonance frequency to reduce the instability threshold has a periodic sawtooth-shaped dependency on the zero-current bunch length.
This periodicity of $\hat{f}_\text{res}$ also makes it possible to effectively influence the beam dynamics with a single impedance structure for several machine settings. The simulation studies reveal that the bursting threshold can be reduced most effectively when the periodicity of the additional wake potential matches the size of substructures in the longitudinal bunch profile.\\
Since adding an impedance enhances or changes the micro-bunching instability's properties, installing a corrugated plate cannot suppress the instability. To reduce or avoid the creation of the substructures, the possibility of lowering the impedance in the frequency range from \SIrange{70}{160}{\giga\hertz} should be investigated.
\begin{acknowledgements}
This work is supported by the DFG project 431704792 in the ANR-DFG collaboration project ULTRASYNC.\\
R.~Ruprecht is acknowledged for his helpful support, knowledge, and contacts for finding suitable manufacturing methods. H.~J.~Cha is acknowledged for his support with the impedance simulations. P.~Schreiber is acknowledged for his support of the inclusion of impedances into Inovesa.\\
S.~Maier acknowledges the support by the Doctoral School "Karlsruhe School of Elementary and Astroparticle Physics: Science and Technology "(KSETA).
\end{acknowledgements}

	
\end{document}